\documentclass{aa}
\usepackage{graphicx}
\usepackage{txfonts}

\begin{document}

\title{
LS~I~+61$^{\circ}$303 in the context of microquasars}

\author{Maria Massi}

\institute{Max-Planck-Institut f\"ur Radioastronomie,
 Auf dem H\"ugel 69, 53121 Bonn, Germany
~(mmassi@mpifr-bonn.mpg.de)}


\date{Received  24 March 2004/ Accepted 27 April 2004}

\abstract{
LS~I~+61$^{\circ}$303 is one of the most observed Be/X-ray binary systems
because,  peculiarly, it has  periodical radio and X-ray
 emission together with  strong,  variable gamma-ray emission.
This source remains,  however, quite enigmatic.
Some properties of this system 
can be   explained assuming that the unseen companion is
 a non-accreting young pulsar with
a  relativistic wind strongly interacting with  the wind of  the Be star.
On the contrary, other properties of LS~I~+61$^{\circ}$303
   fit a model where the   companion  is accreting even with   
 two events of  super-critical accretion  along the orbit.
The very recent discovery  of a radio jet extending ca. 200 AU at
both sides of a central core  has definitely proved the occurrence
of accretion--ejection processes  in this system.
Therefore it is of  great interest to combine  this result
with previous observations  at other wavelengths 
within the framework  of the two-peak   accretion  (ejection) model.
Concerning the first ejection,
we show that the observed  gamma-rays variations   might be  periodic 
with outbursts   confined around the periastron passage
(i.e. where the first accretion-rate peak occurs and high-energy emission
but no radio emission is predicted). 
Concerning the second ejection, with  radio bursts, 
 we point out that  it can be also traced  in the X-ray data, 
both  in episodes of hardening of the X-ray emission 
and  in  a transition from soft- to hard-states at the onsets of radiobursts.
In fact, both hardening and  transitions between spectral states are 
related to the dramatic  change in the structure of the
accretion disk  preceeding the ejection.
Finally, we explore the nature of the accretor and  we conclude 
that on the basis of the present  optical data
a black hole cannot be  ruled out.
}

\maketitle

\section{Introduction} \label{introduction}

The X-ray binary stellar system  LS~I~+61$^{\circ}$303 
has always  excited  particular interest because of its two 
peculiarities: On the one hand being
the probable counterpart of
 the  variable gamma ray source \object{2CG~135+01}
 (Gregory $\&$ Taylor \cite{gregory78}; Tavani et al. \cite{tavani98}) 
and on the other hand  being a periodic radio source
(Taylor \& Gregory \cite{taylor82}, \cite{taylor84}).

The binary system   is composed of a compact object 
in an eccentric orbit around  a
rapidly rotating B0-B0.5 main-sequence star,  
undergoing mass loss through an equatorial disk
 (Hutchings \& Crampton \cite{hutchings81}). 
The orbital period is assumed to be of 26.496 days
determined on a radio data base of over 20 years
(Hutchings \& Crampton \cite{hutchings81}; 
Gregory et~al. \cite{gregory99}; Gregory \cite{gregory02}).

High radio outbursts always peak near phase 0.6,
where  $\Phi$=0  has been set at Julian Date
2443366.775 (Taylor \& Gregory \cite{taylor82}; 
Paredes et al. \cite{paredes90}).
However, lower intensity radio outbursts occur
within the   broad  distribution 
$\Phi_{radio}$=0.45--0.95, due to  
 variations of the equatorial disk in the Be star 
 (Paredes et al. \cite{paredes90};
Gregory et al. \cite{gregory99}; 
 Zamanov \& Mart\'i \cite{zamanov00};
Gregory et al. \cite{gregory02};
Gregory \& Neish \cite{gregory02b}).
From optical and near-infrared observations  the periastron passage is
estimated to be in the
range $\Phi_{periastron}$=0.2--0.5 (Hutchings \& Crampton \cite{hutchings81};
 Mart\'i \& Paredes \cite{marti95}).
One of the fundamental questions concerning the periodic radio
outbursts of \object{LS~I~+61$^{\circ}$303} therefore has been:
Why are the radio outbursts shifted with respect to the periastron passage?

The X-ray emission is also periodic
with period estimates of 
  P= 26.7 $\pm$ 0.2
by  Paredes et al. (\cite{paredes97})
 and    P=26.42$\pm$ 0.05
by  Leahy (\cite{leahy01}), clearly in agreement with
the radio period. 
Quite surprising, however, is  that the X-ray outbursts are offset from the 
radio ones. In fact,
the  two available  simultaneous X-ray and radio 
observations by Taylor et al.(\cite{taylor96})
and by Harrison et al.  (\cite{harrison00}) during an orbital period
show that the  X-ray emission peaks in the phase interval   
$\Phi_{X-ray}$=0.43--0.47 (recalculated  by Gregory \cite{gregory02}) 
while  the radio outburst is offset to that by several days.

Taylor et al.  (\cite{taylor92}) and Mart\'{\i} \& Paredes
(\cite{marti95}) have modelled the
properties of this system in terms of 
 an accretion rate 
$\dot{M} \propto {\rho_{wind}\over v_{rel}^3}$,
(where $\rho_{wind}$ is the density of the Be star wind 
and $v_{rel}$ is the relative speed between the accretor and wind) which 
develops two peaks for  high eccentricities: the highest peak
corresponds to the periastron passage (highest density), 
 while the second peak
occurs when the drop in the relative velocity $v_{rel}$
compensates (because of the inverse cube dependence) the decrease in density. 
From typical parameters of  
 LS~I~+61$^{\circ}$303 derived from near infrared data,
Mart\'{\i} \& Paredes (\cite{marti95}) have shown that both peaks are above
the Eddington limit and therefore one expects 
 that matter is ejected
perpendicular to the plane of the accretion disk.
 Near the periastron the short distance to the Be star enhances
 inverse Compton losses: X-ray
or/and gamma-ray outbursts are expected but no radio bursts.
At the second accretion peak, the
compact object is much farther away from the Be star, so that
the electrons can propagate out of the orbital
plane:  an expanding double
radio source should be observed with a radio interferometer
(Taylor \& Gregory  \cite{taylor84}; Taylor et~al.
\cite{taylor92}; Massi et al. \cite{massi93};  Mart\'{\i} \& Paredes
\cite{marti95}; Gregory et~al. \cite{gregory99}). 

The main problem concerning this model is  that the luminosity 
of \object{LS~I~+61$^{\circ}$303} in the X-ray range
 is only $L_{\rm X}\simeq 10^{35}$~erg~s$^{-1}$ (Maraschi \& Treves
\cite{maraschi81}). 
That is three orders of magnitude lower than the Eddington
limit, even for  a neutron star,
 and in addition the bulk of the energy output seems to be shifted from X-ray to
 $\gamma$-ray wavelengths  with  $L_{\rm \gamma}\simeq10^{37}$erg~s$^{-1}$
(Hartman et~al. \cite{hartman99}).
The difficulty of interpreting these results in the context of 
a super-accretion model has led to an alternative young pulsar model
where a population of  relativistic electrons are
produced at the shock boundary
between the relativistic wind of the young  pulsar and the wind of
the Be star
( Maraschi \& Treves
\cite{maraschi81}; Tavani \cite{tavani95};
Harrison et al. \cite{harrison00}; Hall et al. \cite{hall03}). 
As a matter of fact  such a model  fits the 
time-variable high-energy emission observed near periastron from
the Be/pulsar system PSR B1259-63  (Tavani \& Aaron \cite{tavani97}). 

However,  
the recent discovery  of a radio emitting  jet
extending ca. 200 AU at both sides of a central core
(Massi et al. \cite{massi04})  has shown 
the occurrence of accretion--ejection processes  in 
LS~I~+61$^{\circ}$303.
Therefore, this source seems  very
similar to  the microquasar LS~5039 (Paredes et~al.
\cite{paredes00}) 
also subluminous in the X-ray range (even more than
\object{LS~I~+61$^{\circ}$303}) and also 
having $L_{\gamma} > L_{\rm X}$.
The quite stable gamma-ray emission in that case is explained due
 to upscattered
stellar photons via inverse Compton from the relativistic electrons
 of the persistent  jet. 
If that is true for \object{LS~I~+61$^{\circ}$303}
(Taylor et al.  \cite{taylor96};
 Massi et~al. \cite{massi01};
 Kaufman Bernad\'o et~al. \cite{kaufman02}),
do the periodic outbursts of this source  imply  periodic gamma-ray bursts ?
And in this case at which  $\Phi_{\gamma-ray}$ ? 
Moreover,
recent developments in the theory of the accretion-ejection processes
show that  magneto-rotational instabilities  
are able to accelerate and to collimate a part of the
disk material into a double jet only after thermal 
instabilities in the accretion disk
have inflated and transformed  it into a geometrically thick disk (Meier \cite{meier01}). 
As shown by  Belloni et al. (\cite{belloni97}) 
structural  changes of the disk are associated 
with  changes in the X-ray spectral states.
Is it possible to discern a  change of state in available X-ray
data ?  And in this case: are   changes of state and   
radio outburst related to each other as  expected in an
 accretion-ejection process ? 
The aim of this paper is to try to answer these questions. Section  2
analyses and discusses gamma-ray data while  Section 3 deals with  X-ray and
optical results. The conclusions are given in Section 4.

\section{EGRET data analysis} \label{gamma}

\begin{figure}[htb]
\centering
\resizebox{\hsize}{!}{\includegraphics[scale=0.3, angle=0]{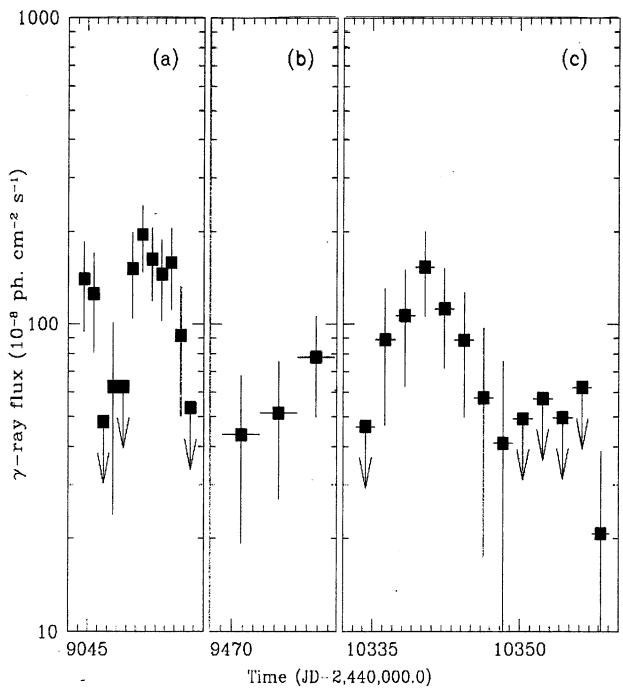}}
\caption{By  Tavani et al. \cite{tavani98}. EGRET light curve  
of  2CG~135+01.}
\label{tavani}
\end{figure}

The EGRET  gamma-ray data  in Fig.~\ref{tavani}, extend over a period of ~1300 days.
During that long interval  there are  three 
 data sets ``a,b,'' and ``c'' lasting for 10, 7 and 22 days, respectively. 
In order to search for eventual periodicitites 
we used the  Phase Dispersion Minimization (PDM) method 
which is 
 very  efficient  on irregularly spaced data (Stellingwerf \cite{stellingwerf78}). 
We used the PDM algorithm  of  the UK Starlink software
package, PERIOD  
(http://www.starlink.rl.ac.uk/).
 
\begin{figure}[htb]
\centering
\resizebox{\hsize}{!}{\includegraphics[scale=0.3, angle=0]{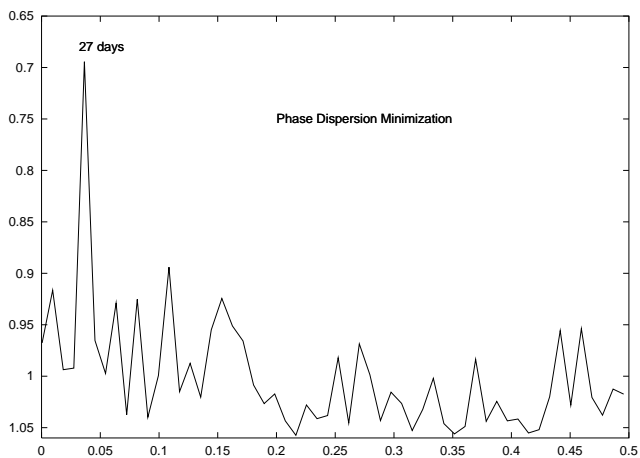}}
\caption{Result by the Phase Dispersion Minimization analysis.  
Note that the analysis
provides the most probable period as a minimum  
 (Stellingwerf \cite{stellingwerf78}) and 
the vertical scale is here shown  reversed. 
 The dominant feature is at P=27.4$\pm$7.2
days.
}
\label{pdm}
\end{figure}

\begin{figure}[htb]
\centering
\resizebox{\hsize}{!}{\includegraphics[scale=0.5, angle=-90]{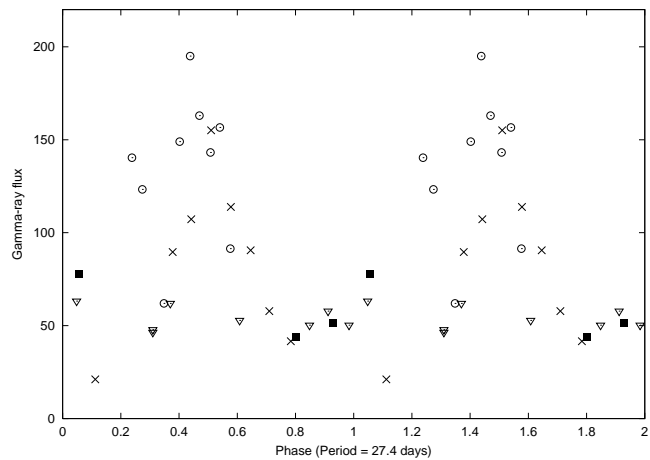}}
\caption{ Gamma-ray observations of Fig.~\ref{tavani}  
folded  with a  period of 27.4 days.
The Phase is computed with respect to  $t_0$= 2443366.775 JD
 (Gregory \cite{gregory02}).  
The circles correspond to the data set "a",
the full squares to the data set "b",
the crosses to the data set "c" and the   
triangles to the upper limits.
}
\label{pdm5}
\end{figure}

As shown in Fig.~\ref{pdm} the  PDM results  in 
a dominant feature at  P=27.4$\pm$7.2.
The folding of the EGRET  data with this period   
results in a  clear clustering of the gamma-ray emission 
(Fig.~\ref{pdm5}).

The statistical significance of the period
is calculated in  PERIOD  following the method of Fisher randomization as
outlined in
 Nemec \& Nemec (\cite{nemec85}).  The advantage of using a Monte-Carlo- or 
 randomization-test is, that it is distribution-free and that it is not constrained
by any specific noise models (Poisson, Gaussian etc.).
The fundamental assumption is:  If there is no periodic signal in the time
series data, then the measured values (gamma-ray flux in our case)
are independent of their observation times and are likely to have occurred in any other order.
One thousand randomized time-series are  
formed and the periodograms calculated. The proportion of
permutations that give   a peak power higher 
(for the PDM: a trough lower) than that of the original time series
would then provide an estimate of p, the probability that no periodic
component is present in the data.
A derived period is defined as significant for p$<$0.01, and a marginally significant  one for 
0.01$<$p$<$0.10 (Nemec \& Nemec \cite{nemec85}).
As compared with other tests, the randomization test is more rigorous in 
rejecting peaks that might contain some oscillatory signal; on the other hand
there is no doubt that the peaks that it finds significant represent an oscillation (Muglach
\cite{muglach03}).
For the period P=27.4 days we got   p=0.009 which implies  an almost zero 
probability that the observed 
time series oscillations could have occurred by chance.
From the extreme case of complete exclusion from
 the data analysis of  the upper limits the result is
that  the period of P=27.4 days is still  determined as a dominant period 
in a noiser periodogram with p=0.099.

\begin{figure}[htb]
\centering
\resizebox{\hsize}{!}{\includegraphics[scale=0.5, angle=-90]{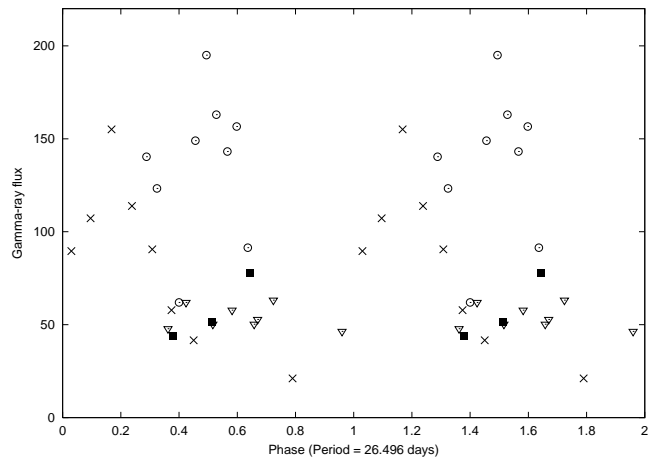}}
\caption{ Gamma-ray observations of Fig. 1 
folded  with the radio  period of 26.496 days.
}
\label{radiophase}
\end{figure}

The gamma-ray emission is predicted to be produced 
via  inverse Compton scattering of 
stellar  photons   by the relativistic electrons of the jet
 at each periastron passage (Taylor et al. \cite{taylor92}, \cite{taylor96}).
Are   the two gamma-ray peaks in the EGRET data indeed 
located around periastron passage ? 
To  determine  the orbital phase of the  gamma-ray emission
we have  folded  the data with the period P=26.496. As shown in 
Fig.~\ref{radiophase}
the two gamma-ray peaks occur at $\Phi_{\gamma-ray}$=0.2--0.5.

\section{X-rays,  and optical observations }

\begin{figure}[htb]
\centering
\resizebox{\hsize}{!}{\includegraphics[scale=0.5, angle=0]{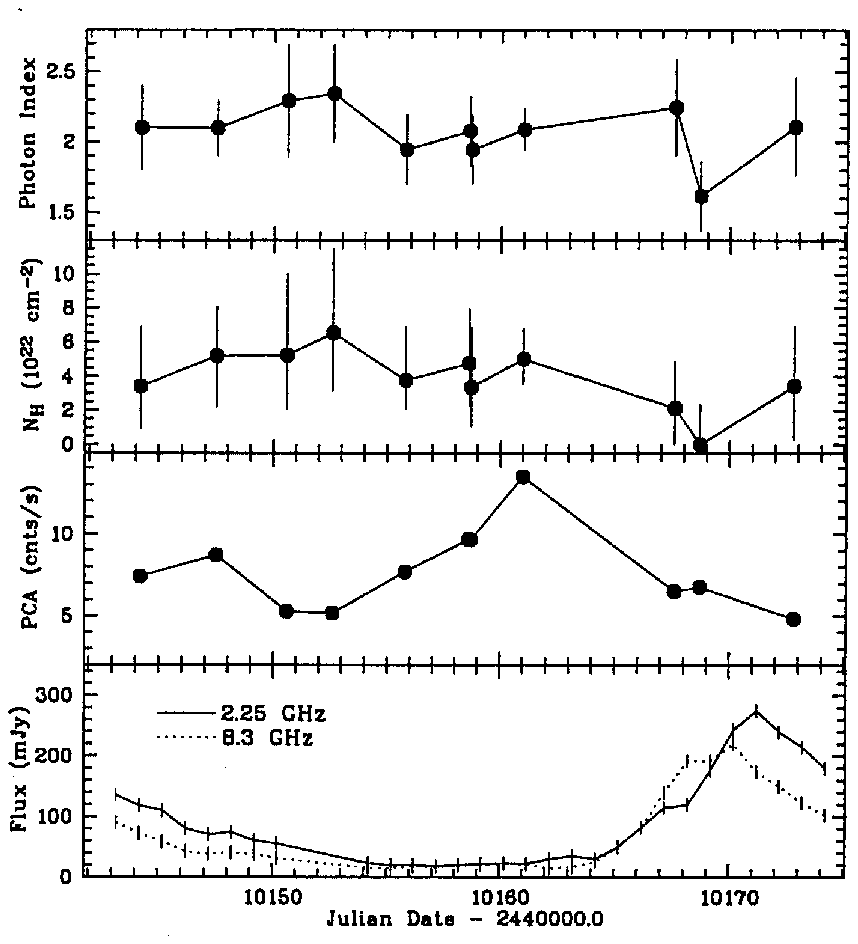}}
\caption{ 
By  Greiner \& Rau  \cite{greiner01}. Top two panels:
 X-ray spectral fit parameters. Bottom two panels: X-rays and radio data. 
}
\label{greiner}
\end{figure}

Variations in the  structure of the accretion disk around a
compact object in a X-ray binary system can  be  revealed  by X-ray data
when they show  a  transition of   the spectral state  
 from the high/soft  state to the low/hard state
(Tanaka \cite{tanaka97}).
 In their high/soft state  systems hosting black holes 
have   a  power-law component E$^{-\Gamma}$  
 (Tanaka \cite{tanaka97}, his Table 3)
with $\Gamma \ge$ 2. 
When the high/soft state evolves into a  low/hard state
the value of $\Gamma$ for the power law becomes  
$\Gamma \sim$ 1.6
(in this case for both X-ray binary systems
containing black holes or neutron stars) (Tanaka \cite{tanaka97}).
The spectral transition from high/soft to low/hard  
corresponds to  a change in the disk structure (Tanaka \cite{tanaka97}).
The radio emission from microquasars with
steady jets  is always correlated with
low/hard states  (Fender \cite{fender04}).

Changes of the hardness ratio  in the source GRS 1915+105  
have shown to 
correspond to  a drastic change of the inner radius of the accretion disk
(Belloni et al. \cite{belloni97}) followed by the  onset of 
synchrotron radiation, first at  infrared wavelengths 
and afterwards at  radio wavelengths
(Mirabel et al. \cite{mirabel98}).

Figure ~\ref{greiner}-top shows the resulting  photon index $\Gamma$ 
of RXTE data of  LS~I~+61$^{\circ}$303   
published by Greiner and Rau (\cite{greiner01}).
As one can see, the range for  $\Gamma$ is 2.0--2.4 
except at one point where the photon index is $\Gamma$=1.6.    
This point at $\Gamma$=1.6 is clearly below the average 
and therefore  assumed to be affected by unknown systematic uncertainties.
In the context of microquasars this value corresponds -
like the  change in the hardness ratio as discussed above - to
a variation in the accretion disk followed by the ejection of matter. And 
indeed  the simultaneous radio observations show    
 (Fig.~\ref{greiner}- Bottom) that the  photon index  value $\Gamma$=1.6    
 coincides with  the onset of a  radio outburst.

The error bars in the photon index are large and  there is  only one
critical point where the transition takes place. However, 
Taylor et al. (\cite{taylor96}) also noticed a 
hardening of the X-ray emission  at the onset
of a new radio outburst. Moreover,   
Leahy et al. (\cite{leahy97}) report on two
X-ray observations, both with power-law indices in the range
1.63--1.90  at orbital phases 0.2 and 0.42.
While for the first value we can only speculate that it
might be coincident with the first accretion rate
peak (no simultaneous gamma-ray observations are available), 
the second value is certainly coincident with the onset of a
radio outburst (Fig. 3 in Harrison et al. \cite{harrison00}).

In conclusion, 
the fit by  Greiner and Rau (\cite{greiner01})
suggests that 1)  LS~I~+61$^{\circ}$303  
with his power-law component   with $\Gamma \ge$ 2 
is mainly in the high/soft  state typical for systems hosting black holes,
and 2) it 
changes this state  into a low/hard state
just before/during (it is difficult to determine the exact time delay)
 the onset of a radio outburst.

The suggestion - coming from the X-ray data -
of a black hole as the accreting object is 
appealing.
Only Punsley (\cite{punsly99})  discussed this possibility in detail
and presented a model for the high-energy emission based on it.
In the literature  it has  generally  been assumed
that  
LS~I~+61$^{\circ}303$ hosts  a neutron star because of an estimated mass of
 M$_x$=1.5 M$\odot$  
by Hutchings \& Crampton (\cite{hutchings81}).
To derive that value Hutchings \& Crampton (\cite{hutchings81})
  assumed a mass function  $f$=0.02,
an inclination i$\simeq$70$^{\circ}$ ($\sin^3$i=0.8)  and a mass for the companion
star of  M=10 M$\odot$. 
Below, we will show that these values are only average values
from a rather large possible range.

Concerning the mass function $f$
as noticed by Punsly  (\cite{punsly99}) the 
observations of  Hutchings \& Crampton (1991)
imply a value in the range  0.0028$<f<$0.043.

Ultraviolet spectroscopy
(Hutchings \& Crampton (\cite{hutchings81})
 indicates that the primary star is a main sequence B0-B0.5 star 
(L$\sim 10^{38}$)
 erg sec$^{-1}$ with a possible range for the mass of
 5--10  M$\odot$. 
Values in the range 10--16.5  M$\odot$ have been suggested 
by Punsly  (\cite{punsly99}) and values in the ranges
10--18  M$\odot$ by Mart\'i $\&$ Paredes
(\cite{marti95}).

 Assuming 600 km s$^{-1}$ as the critical rotational velocity for a
 normal B0 V star and that  Be stars may rotate at a velocity 
not  exceeding 0.9  of this value 
(Hutchings et al. \cite{hutchings79}),
the  very low limit for the  inclination of the orbit compatible
to the measured value  V$\sin$~i = 360 $\pm$ 25 km s$^{-1}$ (Hutchings \& Crampton \cite{hutchings81})
 is of  38$^{\circ}$ (Massi et~al. \cite{massi01}).
On the other hand, 
the  observed (Hutchings \& Crampton \cite{hutchings81})
 shell absorption in the strong Balmer and  He I lines
 for a disk sufficiently flat  corresponds to a large inclination
angle (i=90$^{\circ}$) (Kogure \cite{kogure69}).

In conclusion, the uncertainties of the parameters are  
0.0028$<f<$0.043 for the mass function,
38$^{\circ} <$i$< 90^{\circ}$ for the inclination angle and
 5 M$\odot <$ M$<$ 18 M$\odot$ for the mass. 
The chosen values of $f$=0.02,
 i$\simeq 70^{\circ}$  and M=10 M$\odot$
are indeed  reasonable  average values.
However, already assuming   i=38$^{\circ}$, one obtains   M$_x$=2.5 M$\odot$.
Assuming in  addition  M=18 M$\odot$  the result is  M$_x$=3.4 M$\odot$,
clearly a   black hole. Finally,  
taking $f$=0.043 we are faced with  an  even more  massive object.
We can conclude that on the basis of the present optical  data 
the presence  of a black hole in the system LS~I~+61$^{\circ}303$
cannot be ruled out.  

\section{Conclusions} \label{discussion}

The  emission of  the gamma-ray source 2CG~135+01 is consistent with a
periodic behaviour, 
with a period  similar to  the orbital/radio period  
of the system LS~I~+61$^{\circ}$303. 
The  gamma-ray peaks, observed at two different epochs 
(Tavani et al. \cite{tavani98}),
  remain   confined  around the periastron passage,
a fact which may imply that the most of the seed photons for  Comptonization
are indeed  stellar photons (Taylor et~al. \cite{taylor96}).

The  X-ray observations along an orbital cycle
(Greiner \& Rau \cite{greiner01})
 show that the system remains always
in a  high/soft state, characterized by a photon index $\Gamma \ge 2$.
Only at the onset of
 radio emission there happens a  state transition, characterized by 
 $\Gamma  \sim 1.6$,  to   the
 low/hard state  (Tanaka \cite{tanaka97}, Fender et al. \cite{fender99}).
These  transitions between spectral states are 
related to the dramatic  change in the structure of the
accretion disk  preceeding the ejection of part of it into a jet
(Belloni et al. \cite{belloni97},
 Mirabel et al. \cite{mirabel98}, Fender et al. \cite{fender99}). 
Good sampled simultaneous X-ray and radio observations
in different outbursts would be quite
important  to trace (by monitoring the photon index $\Gamma$ )
the two  accretion-ejection processes occurring  in this peculiar source.

 Finally, we  suggest that LS~I~+61$^{\circ}$30 may host a black hole,
because   its  
 high/soft state has  a power law component with $\Gamma \ge$2,
which  is    
 typical for systems hosting black holes.
On the other hand the value of 
 M$_x$=1.5 M$\odot$ - always quoted in the literature - is 
 related to average values for the  inclination of the orbit, 
mass of the Be star  and mass function.
The ranges for these parameters  presently available are
so large that a massive black hole in LS~I~+61$^{\circ}$303
cannot  be excluded.
 A new determination
 of all these values is desirable.

\begin{acknowledgements}

I am grateful to Karl Menten, J\"urgen Neidh\"ofer,  Guidetta Torricelli,
Leslie Hunt,  Edward Polehampton and  the unknown referee 
for their comments and suggestions.
I acknowledge the data analysis facilities provided by the
Starlink Project which is run by CCLRC on behalf of PPARC and I
am thankful to   Jennifer Hatchell  for the local help.
\end{acknowledgements}

\end{document}